# An experimental test of the local fluctuation theorem in chains of weakly interacting Anosov systems

Giovanni Gallavotti, Fabio Perroni

*Dipartimento di Fisica, Università di Roma I*
*P.le Aldo Moro 2, 00185 Roma, Italia*

**Abstract**

An experimental test of a large fluctuation theorem is performed on a chain of coupled "cat maps". Our interest is focused on the behavior of a subsystem of this chain. A *local* entropy creation rate is defined and we show that the local fluctuation theorem derived in [G1] is experimentally observable already for small subsystems, *i.e.* the finite size effects, for which the theory says little, are small even in small subsystems of the systems considered.

## 1 Introduction

A general law governing the fluctuations of the phase space contraction rate has been derived under various chaoticity assumptions, [GC1], [GC2], [MR], [K], [LS1], [LS2], [M], a review is in [G3], and it has been observed in several experiments, [BGG], [LLP], [CL], [BCL], [RS], [LRB].

Consider a chaotic (*i.e.* Anosov) dynamical system $\mathcal{S} : x \to \mathcal{S}x$, on a phase space $\mathcal{M}$ of "size $\overline{V}$" which we shall call "volume" and which will be the number of interacting subsystems composing our "spatially extended" system, see Sec.2 for a precise definition. Let $\mu$ be the natural stationary state, *i.e.* the SRB distribution, and suppose that $\mathcal{S}$ is a time–reversible dynamics. One can study the fluctuations of the phase space contraction rate, per unit volume, averaged over a time $\tau$:

$$p = \frac{1}{\overline{V}\,\tau\,\overline{\eta}_+} \sum_{j=-\frac{1}{2}\tau}^{\frac{1}{2}\tau} \overline{\eta}(\mathcal{S}^j x), \qquad (1.1)$$

where $\overline{\eta}_+$ is the infinite time average of $\overline{\eta}(x)$: $\overline{\eta}_+ = \lim_{\tau\to\infty} (\overline{V}\,\tau)^{-1} \sum_{j=0}^{\tau-1} \overline{\eta}(\mathcal{S}^j x)$. The value $\overline{\eta}_+$ will be called the "*average phase space contraction*" per unit time and volume. It will also be called the *average entropy creation rate*.

The probability distribution $\overline{\pi}_\tau(p)$ of the fluctuations $p$ in the stationary state verifies, [G1], for large $\tau$:

$$\log \frac{\overline{\pi}_\tau(p)}{\overline{\pi}_\tau(-p)} = \tau\,\overline{V}\overline{\eta}_+\,p + O(\overline{V}) \qquad \text{\textquotedbl fluctuation theorem\textquotedbl}. \qquad (1.2)$$

The size $\overline{V}$ is denoted $O(1)$ in all the papers to which we refer because the authors keep the number of degrees of freedom fixed but here it will be a parameter.

The normalization proportional to $\overline{V}$, used in (1.1), takes into account that in the models that we shall consider dissipation will occur in a translation invariant way so that $\eta$ will have a value typically proportional to $\overline{V}$, hence an average $\overline{V}\overline{\eta}_+$ with $\overline{\eta}_+$ essentially independent of $\overline{V}$, *i.e.* of the spatial dimensions of the system (for $\overline{V}$ large). This implies that we expect, both from a physical and from a mathematical point of view, that in macroscopic systems the observable $p$ has practically *unobservable* fluctuations on the scale involved in (1.2).

*However* if one looks at the motion of a subsystem localized in a spatially microscopic region $V$ and if one considers a *local* entropy creation rate then fluctuations are far more probable. Hence observable consequences of a local fluctuation theorem are likely to become possible.

It is not clear which would be, in any particular case, the appropriate definition of local entropy creation rate, see [BCL]. The simplest situation seems to arise in the case of a chain of $\overline{V}$ ($\overline{V}$ large) very chaotic systems weakly coupled to each other by a translation invariant coupling (see below). For such model systems there is a rather natural definition of local entropy creation, or local phase space contraction: at least it became natural to us after the first few attempts at performing the numerical study that we present in this paper on the basis of other definitions, that in retrospect were not as natural and which failed to give results of any interest.

The definition of local entropy creation rate for the systems that we consider here has been proposed and discussed in [G1], where a local version of the fluctuation law has been derived.

Consider a system with a "large" phase space $\mathcal{M} = \otimes_{j=0}^{\overline{V}-1} \mathcal{M}_0$ where $\mathcal{M}_0$ is the phase space of a "single" system in a *periodic chain* of length $\overline{V}$: hence a point in $\mathcal{M}$ representing a configuration of our chain will be a $\overline{V}$-ple of coordinates $(x_0, \ldots, x_{\overline{V}-1})$. Let $V$ be a small *fixed* interval $[0, V-1] \subset [0, \overline{V}-1]$, and let $\eta(x)$ be the "local phase space contraction rate" (yet to be defined and in general different from the $\overline{\eta}(x)$ above) in the subsystem described by the coordinates $(x_0, \ldots, x_{V-1})$. Define $\eta_+ = \lim_{\tau \to \infty} \frac{1}{V\tau} \sum_{j=0}^{\tau-1} \eta(\mathcal{S}^j x)$, and extend the definition (1.2) above to $V \leq \overline{V}$ as

$$p = \frac{1}{V\tau\eta_+} \sum_{j=-\frac{1}{2}\tau}^{\frac{1}{2}\tau} \eta(\mathcal{S}^j x). \tag{1.3}$$

Then, if $\pi_\tau(p)$ denotes the probability distribution of the above $p$, it verifies, see [G1]:

$$\log \frac{\pi_\tau(p)}{\pi_\tau(-p)} = \tau V \eta_+ p + O(\tau + V). \tag{1.4}$$

where one should note that $\tau + V$ is (half) the size of the boundary of the space–time region involved in the definition of $p$ in (1.3). When $V = \overline{V}$ the error is only $O(\overline{V})$ rather than $O(\tau + \overline{V})$, see (1.2), because there is no "vertical" boundary and, therefore, no boundary effect related to it, see [G1]. In (1.4) and in the cases that we shall consider, $\eta_+$ can be replaced by $\overline{\eta}_+$ because it will be $\eta_+ = \overline{\eta}_+ + O(V^{-1})$.

The relations (1.2), (1.4) are the objects of our experimental test in a concrete, simple, model for $\mathcal{M}_0$ and $\mathcal{S}$.

One can ask "why to make the test if there is a theorem that predicts the results?" This is true: but the theorem only leads to estimates of the remainders in (1.2), (1.4). Such estimates are not quantitatively derived in the proof, because various constants in the bounds are only shown to be finite with no attempt at computing them. They can be, however, computed following a rather general approach (see [G1] where the general theory of [PS], [BK1], [BK2], [JP] is used in an essential way) and, as usual when applying general mathematical theorems to concrete cases, we expect the estimates to be very poor.

Therefore it makes sense performing an experiment to check if the predictions of the theorem *are actually observable*: a situation similar to the one met in the theory of phase transitions and in the numerical experiments performed on the two dimensional Ising model. The latter model is exactly soluble in the sense that many of its properties can be analytically computed: nevertheless experiments on its *a priori known* properties make sense and are important to test the feasibility and the reliability of simulations on systems with an enormous number of degrees of freedom and to test the accuracy and reliability of methods that one really wants to apply to more complicated systems not amenable to an exact solution.

The global fluctuation laws (1.2) are universal in many systems where, however, they are extremely difficult to observe: in this paper we give evidence that their local counterparts can be observed quite easily because they manifest themselves observably already in rather small subsystems.

This is quite analogous to the theory of density fluctuation in a gas: one cannot observe density fluctuations on large gas samples because their likelihood decreases exponentially with the volume. Nevertheless such fluctuations are of great interest (their value will give us the compressibility of the gas) and they can be measured by observing them in small samples and by making use of the fact that the logarithms of the distribution of their averages over a given volume $V$ are proportional to it so that once the fluctuations are known in a small sample one can infer their (otherwise practically unobservable) probability in a large sample.

In section 2 we describe the particular system considered in our experiment and we define the local phase space contraction rate following [G1]; in section 3 we expose the experimental results of our simulations. A few comments will follow in section 4. An appendix discusses briefly our method to estimate the experimental errors that are characteristic of our simulations.

## 2 Description of the system

Let $M'$ be a two-dimensional torus, $M' = [0,1] \times [0,1]$, and $\underline{x}$ a pair of coordinates on it. A *Arnold cat map* $S' : M' \to M'$ is the dynamics on $M'$ generating from an initial datum $\underline{x} = \underline{x}^0$ a trajectory $t \to \underline{x}^t$

$$\underline{x}^{t+1} = S\underline{x}^t = \begin{pmatrix} 1 & 1 \\ 1 & 2 \end{pmatrix} \underline{x}^t \mod 1. \tag{2.1}$$

Consider a finite number $\overline{V}$ of copies of such systems and label each copy with an index $i \in \{0, \ldots, \overline{V} - 1\}$ identifying $\underline{x}_0$ and $\underline{x}_{\overline{V}}$ (*periodic boundary conditions*).

Let $\mathcal{M}' = \bigotimes_{i \in \overline{V}} M'_i$ be the *phase space* of such new "extended" system, let $\underline{x} = \bigotimes_{i \in \overline{V}} \underline{x}_i$ be a point in it and let $\mathcal{S}' = \bigotimes_{i \in \overline{V}} S_i$ its ("unperturbed") evolution: this means that each coordinate $x_i$ evolves under the map $\mathcal{S}'$ into $S_i x_i = (\mathcal{S}' x)_i$. We now consider a perturbation of the map $\mathcal{S}'$

$$\widetilde{\mathcal{S}}(\underline{x}) = \mathcal{S}'(\underline{x}) + \varepsilon \mathcal{F}(\underline{x}) \tag{2.2}$$

where $\mathcal{F}$ is a "nearest neighbor coupling" so defined that:

$$\widetilde{\mathcal{S}}(\underline{x})_i = \mathcal{S}'_i(\underline{x}_i) + \varepsilon \underline{f}_i(\underline{x}_{i-1}, \underline{x}_i, \underline{x}_{i+1}). \tag{2.3}$$

We choose $\underline{f}_i = (f_i, 0)$ with $f_i$ very simple. Namely, if $\underline{x}_j = (\zeta_j, \xi_j)$, for $j = 0, \ldots, \overline{V} - 1$, and if we set $\alpha_j = 2\pi(\xi_{j-1} + \xi_j + \xi_{j+1})$ and $\beta_j = 2\pi(\zeta_{j-1} + \zeta_j + \zeta_{j+1})$:

$$f_i = \frac{1}{2\pi}[\sin(\alpha_i + \beta_i) + \sin \beta_i] \tag{2.4}$$

On the boundaries we simply set $\zeta_{-1} = \zeta_{\overline{V}-1}$, $\xi_{-1} = \xi_{\overline{V}-1}$, $\zeta_{\overline{V}} = \zeta_0$, $\xi_{\overline{V}} = \xi_0$, *i.e.* periodic boundary conditions.

The system $(\mathcal{M}', \widetilde{\mathcal{S}})$ is, for small perturbations (*i.e.* small $\varepsilon$ but for all $\overline{V}$), an Anosov map by the remarkable extension of structural stability to chains of Anosov maps with short range interaction developed in [PS]. However the theory of the fluctuations that we test here applies directly only to "*reversible* systems, see [GC1, GC2] and below.

A map $\mathcal{S}$ is a *time reversible map* for a *time reversal operation* $\mathcal{I}$ if $\mathcal{I}$ is a smooth isometric involution $\mathcal{I} : \mathcal{M} \longleftrightarrow \mathcal{M}$, $\mathcal{I}^2 = Identity$, such that $\mathcal{I}\mathcal{S} = \mathcal{S}^{-1}\mathcal{I}$.

Therefore the system that we shall consider in the experiment cannot just be the one described so far because the map $\mathcal{S}$ is not time reversible (at least not for any simple map $\mathcal{I}$). We shall, however, consider a closely related system which has a *time–reversal* symmetry and which we now describe.

Let $\mathcal{M} = \mathcal{M}' \otimes \mathcal{M}'$ and define:

$$\mathcal{S}(\underline{x}, \underline{y}) = (\widetilde{\mathcal{S}}(\underline{x}), \widetilde{\mathcal{S}}^{-1}(\underline{y})), \tag{2.5}$$

The system $(\mathcal{M}, \mathcal{S})$ is (trivially) *reversible* because the map $\mathcal{I}(\underline{x}, \underline{y}) = (\underline{y}, \underline{x})$ is a time reversal for it (see [G2]).

Denote $x = (\underline{x}, \underline{y})$ a point in the phase space $\mathcal{M}$: the "*phase space contraction rate*" of $(\mathcal{M}, \mathcal{S})$ is defined, see [GC1], in terms of the determinant $\Lambda_{\overline{V}}(x) = |\det \partial \mathcal{S}(x)|$ of the jacobian $\partial \mathcal{S}(x)$ of the map $\mathcal{S}$ and of the determinant $\widetilde{\Lambda}_{\overline{V}}(\underline{x}) = |\det \partial \widetilde{\mathcal{S}}(\underline{x})|$ of the map $\widetilde{\mathcal{S}}$.

The *global entropy creation rate* for this system is defined as minus the logarithm of the Jacobian determinant of $\mathcal{S}$, $\Lambda(x) = -|\det \partial \mathcal{S}(x)|$, [GC2], $\overline{\eta}(x) = -\log \Lambda(x)$:

$$\begin{aligned}
\overline{\eta}_{\overline{V}}(x) &= -\log \Lambda_{\overline{V}}(x) = -\log |\det \partial \mathcal{S}(x)| \\
&= -\log \left| \det \begin{pmatrix} \partial \widetilde{\mathcal{S}}(\underline{x}) & 0 \\ 0 & \left(\partial \widetilde{\mathcal{S}}(\widetilde{\mathcal{S}}^{-1} \underline{y})\right)^{-1} \end{pmatrix} \right| \\
&= -\log \left( \widetilde{\Lambda}_{\overline{V}}(\underline{x}) \left(\widetilde{\Lambda}_{\overline{V}}(\widetilde{\mathcal{S}} \underline{y})\right)^{-1} \right)
\end{aligned} \tag{2.6}$$

If $\mu_\pm$ are the forward and backward statistics of the volume measure (*i.e.* the SRB distributions for $\mathcal{S}$ and $\mathcal{S}^{-1}$) the perturbation makes, for generic perturbations $f$, see [BGM], the system $(\mathcal{M}, \mathcal{S})$ dissipative, in the sense that:

$$-\int_{\mathcal{M}} \mu_\pm(d\underline{x}) \log \Lambda_{\overline{V}}^{\pm 1}(\underline{x}) = \overline{\eta}_\pm \overline{V} > 0. \tag{2.7}$$

at least for $\varepsilon \neq 0$ small. This is a property that we checked in our particular case (the form in 2.4 was so chosen to make the algebraic part of this check simpler) and which is a quite general fact, see [R].

Note that time reversal symmetry implies that $\overline{\eta}_+ = \overline{\eta}_-$ and $\overline{\eta}_\pm = 2\widetilde{\eta}$ if $\widetilde{\eta}\overline{V}$ is the average of $-\log \widetilde{\Lambda}_{\overline{V}}(\underline{x})$.

Fixed a subvolume $V$, *i.e.* a subinterval of $\overline{V}$, let $\partial \mathcal{S}_V(x)$ be a block matrix of the Jacobian $\partial \mathcal{S}(x)$ corresponding to the coordinates $\underline{x}_V$ of this subsystem. Set $\widetilde{\Lambda}_V(x) = |\det \partial \mathcal{S}_V(x)|$ and define the *local phase space contraction rate*, see [G1], as

$$\eta_V(x) = -\log \left( \widetilde{\Lambda}_V(\underline{x}) \left( \widetilde{\Lambda}_V(\widetilde{\mathcal{S}}^{-1}\underline{y}) \right)^{-1} \right) \tag{2.8}$$

that will also be called the *local entropy creation rate* in the volume $V$ (we imagine $V \ll \overline{V}$). We will study the fluctuations of the averages over a time $\tau$ of $\overline{\eta}_{\overline{V}}(\underline{x})$ and of $\eta_V(\underline{x})$, see (1.1), (1.3).

## 3  Experimental results

We considered chains of $\overline{V} = 8, 16, 32, 64$ maps as defined in Sec.2, and for each $\overline{V}$ subsystems of size $V = 1, 3, 6$ consisting of $V$ consecutive elements of the chain. The perturbation parameter $\varepsilon$ was given the values 0.05, 0.10, 0.15 but the complete tests were done only for the value 0.10 and the other values were used only for the purpose of checking/testing the codes.

According to the definition in Sec. 2 we associate with each chain site two pairs of coordinates each defining a point on the 2–torus: the collection of the first $\overline{V}$ pairs evolves under the action of a forward map $\widetilde{\mathcal{S}}$, while the collection of the second pairs evolves under $\widetilde{\mathcal{S}}^{-1}$. The inversion of $\widetilde{\mathcal{S}}$ has to be done numerically, because although the map $\widetilde{\mathcal{S}}$ is very simple, its inverse cannot be elementarily expressed: we used a general Newton's algorithm (called Newton–Raphson algorithm in [NR]) which strongly limits the maximal size of the perturbation $\varepsilon$ but which, for the relatively small values of $\overline{V}$ we considered, is competitive with the $O(\overline{V})$-size algorithm that we designed specifically for our problem.

The total number of degrees of freedom is $4\overline{V}$, hence it went up to 256 although the results that we report deal with the $\overline{V} = 16$, *i.e.* with 64 degrees of freedom cases.

For such values of $\varepsilon$ the Lyapunov exponents can be checked to be still very close to the ones corresponding to the case $\varepsilon = 0$. Just as an example we give in Figure 1 the graph of the 64 Lyapunov exponents corresponding to a model with $\overline{V} = 16$ and $\varepsilon = 0.10$.

Note that the intrinsic complexity of the inversion of the map $\widetilde{\mathcal{S}}$ *is not the only obstacle* to the inversion of the map: as $\varepsilon$ increases the system becomes more and more dissipative and at some critical value stops being conjugated to the unperturbed system. For instance the attractor ceases to be dense in phase space, and density is a condition for the validity of the fluctuation theorem, see [GC2]. Although this new situation is interesting we shall not study it here, see [BG].

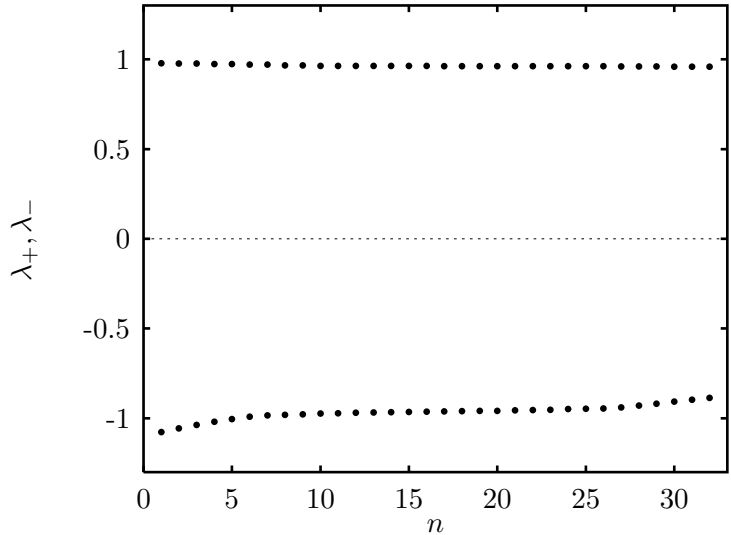

Figure 1: Plot of the Lyapunov exponents for the system with $\overline{V} = 16$, $\varepsilon = 0.10$. The 64 exponents (each "cat" has 2 components that evolve forward and 2 that evolve backwards in time) are ordered in couples $\lambda_+, \lambda_-$ coupling the greatest positive with the greatest negative, the second greatest positive and the second greatest negative and so on.

Our simulations were carried out over $4 \cdot 10^6 \div 300 \cdot 10^6$ iterations[1], depending on the system dimension and on the perturbation parameter $\varepsilon$[2].

The figures 2, 3, 4, 5 refer to the analysis of data produced by the system with $\overline{V} = 16$, $\varepsilon = 0.10$ for $N = 3 \cdot 10^8$ iterations. The values of $\overline{V} = 8, 32, 64$ as well as $\varepsilon = 0.05, 0.15$ have also been considered and we do not report the results. We only say that they were very similar, with the obvious variation that in the cases $\overline{V} = 32, 64$ substantially less statistics was available for the analysis of the global fluctuations (which become too rare as the size of the system increases).

Our data and error analysis follows from the one described and applied in [BGG], see the appendix below for further details.

Once the values of $\varepsilon$, $\overline{V}$ and $V \leq \overline{V}$ are fixed we measure the fluctuation of entropy creation rate $p$ (see (1.1) or (1.3)) averaged over the iteration time $\tau$ ranging from 2 to 400, discarding $\tau_0$ iterations between each measure to let them decorrelate. We set $\tau_0 = 10$ because we know from the computation of the Lyapunov exponents, reported in Figure 1, that the maximal and the minimal ones are such that $\tau_0 = 10$ is about ten times the characteristic time of the system[3]. For the same reason we discard the first 10 iterations in order to reduce the effects of transients while the motion approaches the attractor.

---

[1] To let each simulation take few days of CPU time on a modern PC.

[2] The number of steps the algorithm takes to calculate the backward evolution depends on $\varepsilon$: *i.e.* for $\varepsilon = 0.05$ the algorithm converges (in the sense that the precision reaches the numerical double precision accuracy) in 5 or 6 steps while for $\varepsilon = 0.15$ we have to use a separate algorithm to search for a "good" starting point for the Newton algorithm to converge in 50 steps.

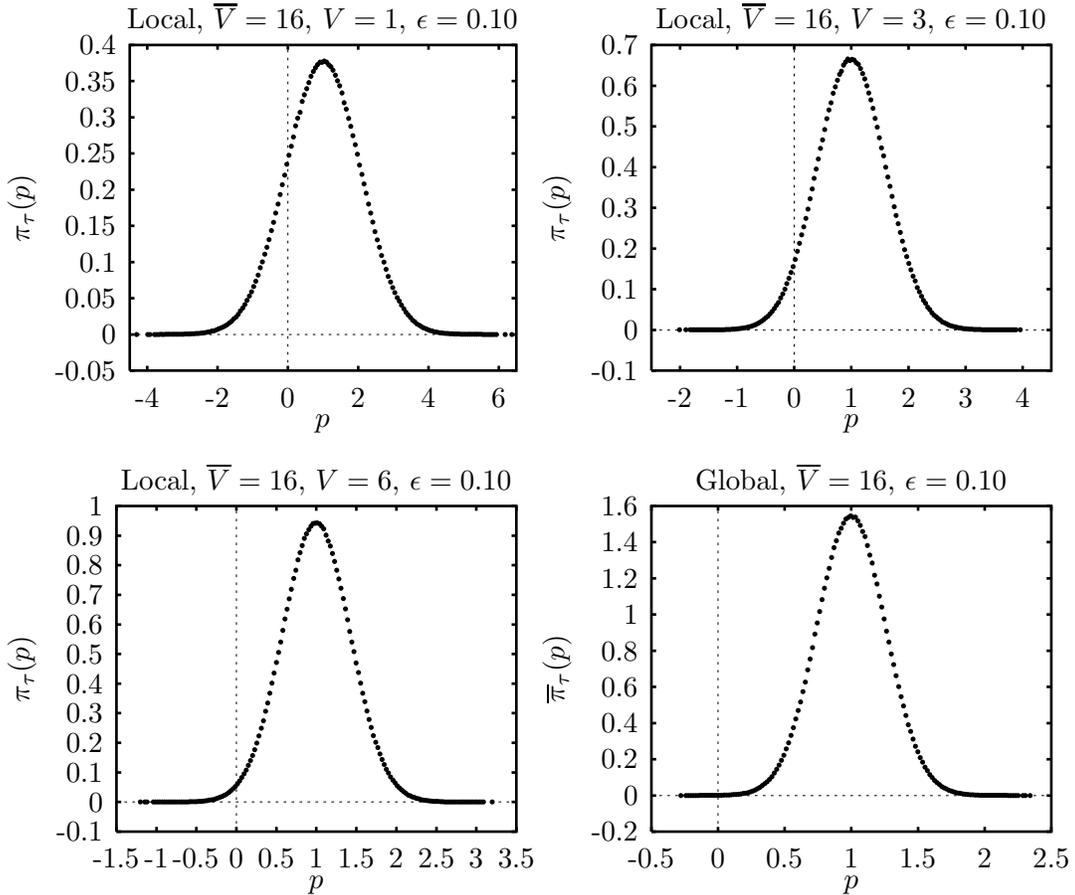

Figure 2: Histogram of the distribution $\pi(p)$ for the chain of $\overline{V} = 16$ subsystems and sub–chains of $V = 1, 3, 6$ with $\varepsilon = 0.10$ and $\tau = 40$. The simulation was carried over $N = 10^8$ iterations. The amplitude of the distribution increases in smaller subsystems: large fluctuations are more probable and the fluctuation theorem can be easily observed. The probability of observing negative values of $p$ in the full system is so small that we could obtain just few experimental points. The error bars (barely visible in the scale of the figure) are discussed in the appendix.

Then we compute the empirical probability distribution of fluctuations of the phase space contraction rate $\overline{\pi}_\tau(p)$ for the whole chain and $\pi_\tau(p)$ for the subsystem, Figure 2.

---

[3]The exponents are of the same order of the unperturbed positive exponents that are exactly $\log\left(\frac{1}{2}(3 + \sqrt{5})\right) \sim 0.98$.

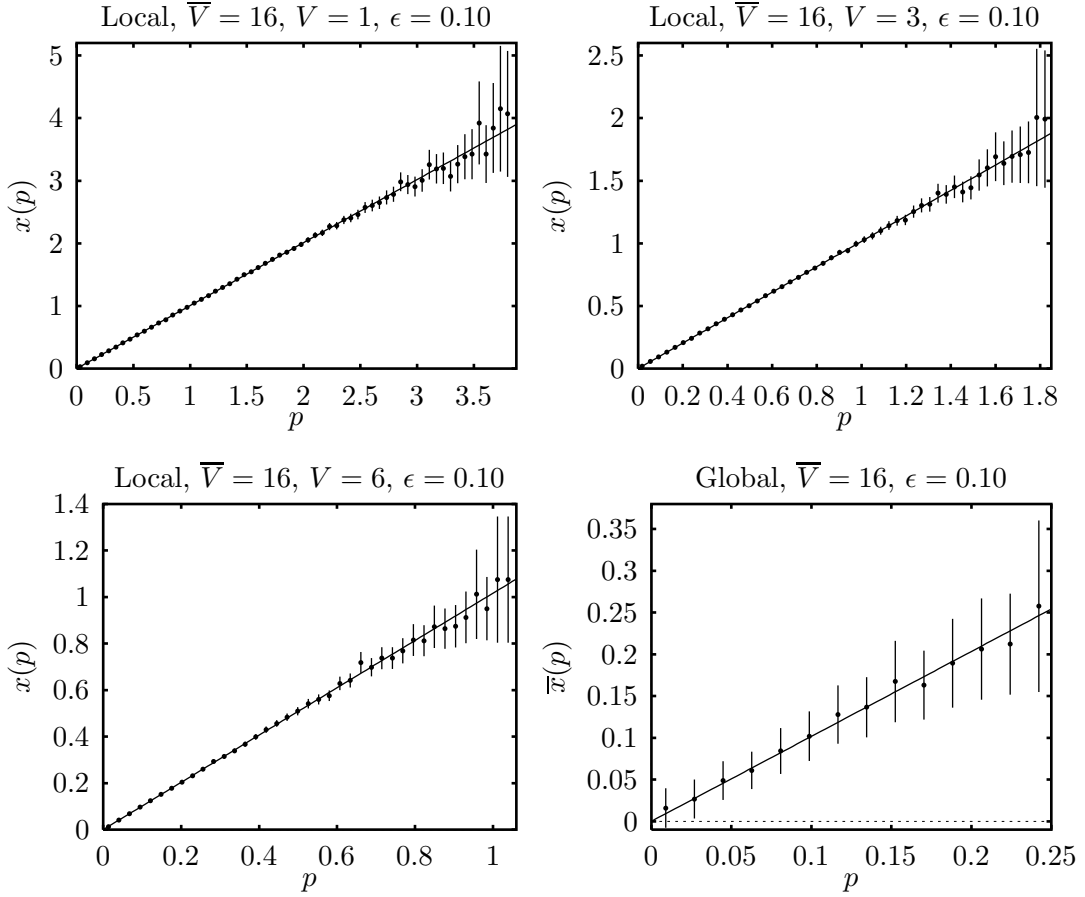

Figure 3: Plot of $x(p)$, (3.1), for the system with $\overline{V} = 16$, $\varepsilon = 0.10$, $V = 1, 3, 6$ and $\tau = 40$. The solid line is the linear least–square fit; the dashed line is the line with slope 1. In smaller subsystems large deviations are more probable (the tails of $\pi_\tau(p)$, Figure 2, go far below 0) and the fluctuation theorem prediction for $\tau = +\infty$ can be observed more easily than in the full system.

To check the predictions of the global and local fluctuation theorems, (1.2) and (1.4), we verify that the quantities

$$\overline{x}(p) = \frac{1}{\overline{\eta}_+ \overline{V}\tau} \log \frac{\overline{\pi}_\tau(p)}{\overline{\pi}_\tau(-p)}, \qquad x(p) = \frac{1}{\eta_+ V\tau} \log \frac{\pi_\tau(p)}{\pi_\tau(-p)} \qquad (3.1)$$

are linear in $p$, see Figure 3, with slope $\chi_\tau$:

$$x(p) = \chi_\tau p, \qquad \chi_\tau \to 1 \quad \text{for} \quad \tau \to \infty \qquad (3.2)$$

We computed $\chi_\tau$ with a linear least squares fit. Results, plotted in Figure 4, show that $\chi_\tau$ does not contain 1 within the error bars only for small values of $\tau$. We attribute this to finite–size effects, [BCL], *i.e.* as a manifestation of the $O(\tau^{-1})$ and, respectively, $O(\tau^{-1} + V^{-1})$ of the corrections in (1.2), (1.4). We also checked that $\eta_+ = \overline{\eta}_+ + O(V^{-1})$ and in fact the correction in $O(V^{-1})$ is not detectable for $V \geq 16$.

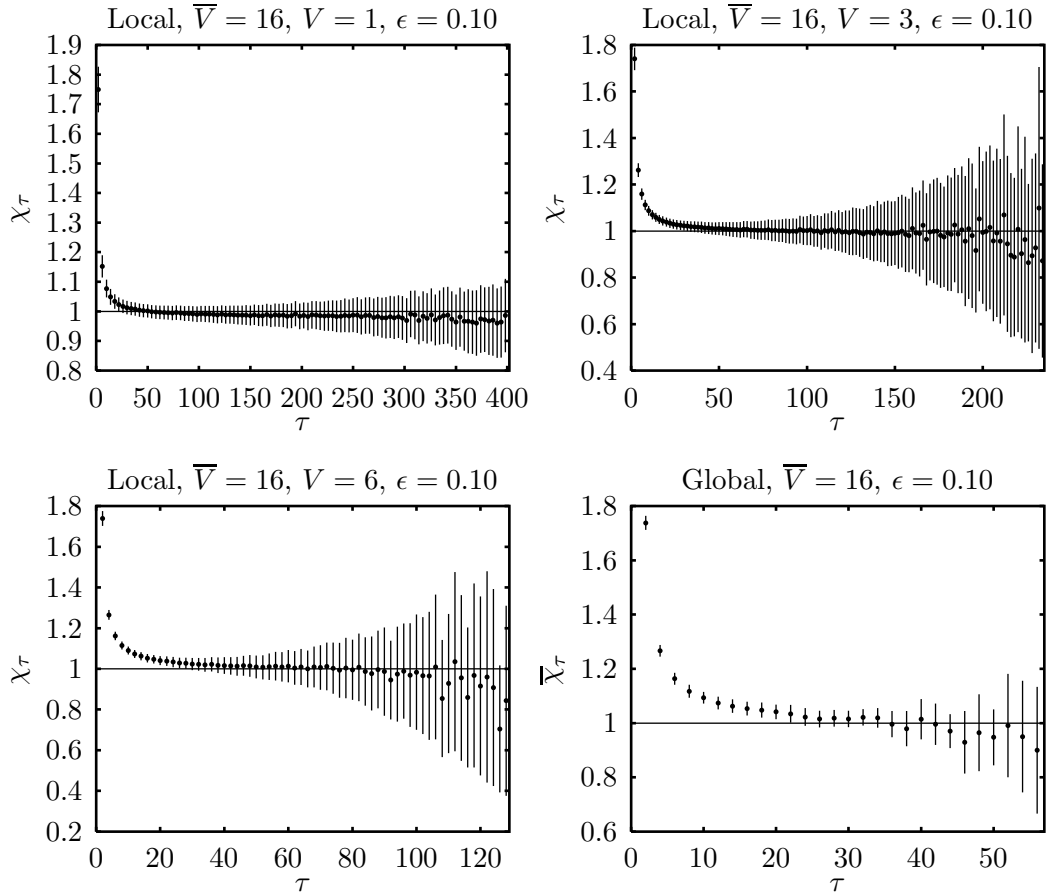

Figure 4: Plot of the slope, $\chi_\tau$, of $x(p)$ for the system with $\overline{V}=16$, $\varepsilon=0.10$ and $V_0=1,3,6$. In smaller subsystems the fluctuation theorem holds for larger ranges of values of $\tau$. For $\tau$ small, $\chi_\tau$ does not contain 1 within the error bars: we attribute this to finite-size (of $\tau$) effects. The straight line has the theoretical slope 1.

A least square fit of the probability distributions $\overline{\pi}_\tau(p)$ and $\pi_\tau(p)$ gives a *Gaussian as best fit* (in the sense of [BGG], [GG]).

If $p$ was Gaussian then it would also verify (3.2) with some slope (not necessarily 1): *however there is no reason for $p$ to be a Gaussian variable*, and in fact it can only be such for $p$ rather close to its average value $p=1$. In fact around $p=1$ it looks Gaussian because of a central limit theorem generally valid for the SRB distribution of Anosov systems, see [S], or of weakly interacting chains of such systems, see [PS], [BK1], [BK2], [JP].

To test whether $p$ can be really taken Gaussian, we computed the second moments of the distributions $\overline{\pi}_\tau(p)$ and $\pi_\tau(p)$. Assuming that $\overline{\pi}_\tau(p)$ and $\pi_\tau(p)$ are Gaussian and agree with the central limit theorem (which, *however*, only really holds for $|p-1| < O((\tau V)^{-\frac{1}{2}})$ implies that their $2^{nd}$ moments $\overline{m}_2(\tau)$ and $m_2(\tau)$ should be linear in $(\tau V)^{-1}$ (to leading order in $\tau V$) and we would denote them $\overline{m}_2(\tau) = (\tau \overline{V})^{-1}\overline{A}$,

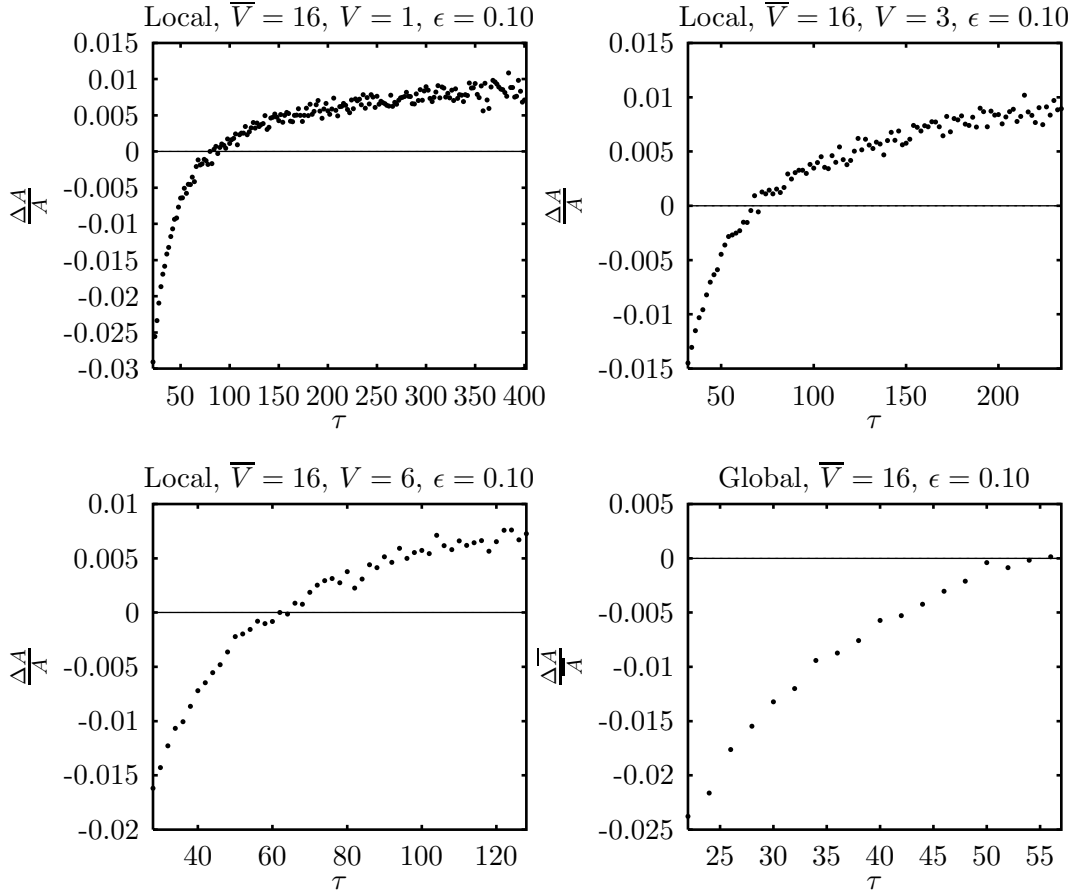

Figure 5: Plots of $\frac{\Delta A}{A}$ for the system with $\overline{V} = 16$, $\varepsilon = 0.10$ and $V_0 = 1, 3, 6$. The experimental values of $\Delta A/A$ differ from 0. The experimental value of $A$ differs from the value $A_0$ that would follow from a Gaussian assumption combined with the fluctuation theorem by $\sim 1\%$, quite clearly visible beyond the error bars. The range of $\tau$ in the graphs is over the values of $\tau$ corresponding to $\chi_\tau = 1$, see Figure 4.

$m_2(\tau) = (\tau\, V)^{-1}\, A$. The distributions of $p$ would then be proportional to:

$$e^{-\tau \overline{V}(p-1)^2/(2\overline{A})}, \qquad e^{-\tau V(p-1)^2/(2A)} \qquad (3.3)$$

for some $\overline{A}, A$ respectively.

Therefore the fluctuation relations (1.2), (1.4) or (3.2) will imply that there should be a simple relation between the slopes 1 of $x(p)$ and $\overline{x}(p)$ and the mean square deviations coefficients $\overline{m}_2(\tau) = (\tau\, V)^{-1}\overline{A}$, $m_2(\tau) = (\tau\, V)^{-1} A$ of the random variable $(p-1)$, namely:

$$\frac{2}{\overline{\eta}_+} = \overline{A}, \qquad \frac{2}{\eta_+} = A. \qquad (3.4)$$

Calling $\overline{A}_0, A_0$ the values in (3.4) we deduce from the experimental of $\overline{m}_2(\tau)$ and $m_2(\tau)$ as functions of $\overline{V}\tau$ or of $V\tau$, the actual experimental values $\overline{A}$ and $A$ (performing a least square fit). We then compare these values with the ones predicted in (3.4) (*i.e.* with $\overline{A}_0 = 2\overline{\eta}_+^{-1}$, $A_0 = 2\eta_+^{-1}$).

$$\frac{\Delta \overline{A}}{A} = \frac{\overline{A} - \overline{A}_0}{\overline{A}_0}, \qquad \frac{\Delta A}{A} = \frac{A - A_0}{A}. \tag{3.5}$$

In Figure 5 we plot $\Delta A/A$ only, of course, *for the values of $\tau$ for which the fluctuation theorem can be checked* (*i.e.* for the values of $\tau$ for which there is enough statistics to have significant results in spite of the error bars, see Figure 4). The results of this test is plotted in Figure 5: The amplitude of the error bars doesn't let us see that the fluctuations of the entropy production are not gaussian (in a numerical test sense) although the value of $\frac{\Delta \overline{A}}{A}$ is not zero. The error analysis seems to play a fundamental role in this test.

## 4  Conclusions

**(1)** The "fluctuation theorem", for our models, is an *exact* result only *asymptotically* when $\tau \to \infty$ in the global case and when $\tau, V \to \infty$ in the local case. It appears, nevertheless, to be observable already for small values of $\tau, V$ provided the "correct" definition of local entropy creation rate is applied (see (2.7), finding it was one of the starting points of the present work, see [G1]). The observability of the predictions of the theorem on small systems is evident for $\overline{V} = 16$ as can be seen in Figure 2, 3, 4. We have performed similar experiments for systems with $\overline{V} = 8, 32, 64$ obtaining quantitatively similar results with (of course) lower statistics in the last two cases: we do not report them for brevity. We also tested, although not in the same detail and the same accuracy, the values of $\varepsilon = 0.05$ and $\varepsilon = 0.15$, the latter being substantially more time consuming: the results seemed to agree with the ones that we report.

**(2)** Large fluctuations are practically unobservable in large systems: it's evident also in the cases we considered. The plot of the probability distributions $\overline{\pi}_\tau(p)$ of the fluctuations of $p$ in the whole chain has just few points for negative values of $p$ (Figure 2) thus the fluctuation theorem can be checked for few values of $\tau$ (Figure 4). However the *local* fluctuation theorem holds for any small subsystem and can be checked for a larger number of values of $\tau$ (Figure 4) *even when the global theorem can not be verified.*

**(3)** The hypothesis of $p$ being Gaussian for either the whole chain or for its subsystems is theoretically "impossible" because large deviations of normalized sums of independent nongaussian variables (*i.e.* of averages of independent equally distributed nongaussian random variables) are not Gaussian *even* in the most random systems like Bernoulli schemes, [FGP].

If the distribution of $p$ was a Gaussian centered at $p = 1$ the second moment of $p - 1$ should have the form $(V\tau)^{-1}A$ and the slope of the logarithm of the odd (in $p$) part of the distribution of $p$ should have the form $2V\tau/A$ and *it should be $A_0 = A$*, which can be interpreted as the validity of the Green–Kubo formula in the nonlinear regime we are considering, see [ES] and Sec. 5 in [BGG] and [G3], [G4].

We plan an experiment to investigate more accurately the results of Sec. 3 indicating that $\frac{\Delta A}{A}$ is significantly nonzero: this means that we are trying to find systems for which the numerical construction of the trajectories is less time consuming for the values of $\overline{V}$ that we consider.

The nongaussian nature of the fluctuations has already been studied and clearly shown in the work [ES] at least in models out of equilibrium. Non gaussian behavior has been also studied and announced by [G]. We find it also in equilibrium (a result not reported here) as (of course) expected.

**(4)** The above considerations are also supported by the fact that the nongaussian nature of the fluctuations is clearly demonstrated in a situation in which, nevertheless, the linearity of the odd part of the logarithm of the distribution of $p$ can be observed (although it is not clear whether it falls within the theory of reversible Anosov systems): this is a case quite different from the presently considered chains of Anosov systems, see [LLP], and the nongaussian nature of $\zeta(p)$ is there manifest for $\tau$ relatively small because the distribution appears sensibly non symmetric around $p=1$.

## A  Appendix: Error evaluation.

We follow the error estimates described in [GG], [BGG]; we add here explicitly the description of the technique we use to estimate errors, which are not trivially due to statistical errors, on the distribution histogram, because in the present experiment it is the main error source.

Let $I(k)$ be the count of the number of values of $p$ in the $k$-th interval $[kd, kd+d]$ (with $d>0$ a fixed mesh) into which the $p$–axis is divided to perform the histogram, and let $I'(k) = I(-k-1)$. The main source of errors is that $\eta_+$ itself is affected by an error; the histogram error due to an uncertainty $\delta_\eta$ in $\eta_+$ is the error in the number count $I(k)$ over the interval $[kd, kd+d]$ given by $2\frac{p}{d}\frac{\delta_\eta}{\eta_+}I(k)$: because the error $\frac{\delta p}{p} = \frac{\delta\eta}{\eta_+}$ in the measurement of $p$ amounts to an uncertainty $p\frac{\delta\eta}{\eta}$ on the apparent position of the *two* extremes of the interval $[kd, kd+d]$ centered at $p = (k+\frac{1}{2})d$.

So that, adding the statistical error $3\sqrt{I(k)}$, the total error on $I(k)$ is:

$$\Delta I(k) = 3\sqrt{I(k)} + 2\,I(k)\,\frac{p}{d}\frac{\delta_\eta}{\eta_+}d \tag{A.1}$$

The corresponding variation of the observed value of $x$:

$$x = \frac{1}{\eta_+}\frac{1}{\tau V}\log\frac{I(k)}{I'(k)} \tag{A.2}$$

has to be taken (recalling that the ideal result is $x(p) = p = (k+\frac{1}{2})d$):

$$\delta_x = \Delta\left(\frac{1}{\eta_+\,\tau V}\log\frac{I(k)}{I'(k)}\right) = p\left(\frac{\delta\eta}{\eta_+} + \frac{3}{\sqrt{I}} + \frac{3}{\sqrt{I'}} + 2\,p\,\frac{\delta_\eta}{d\eta_+} + 2\,p\,\frac{\delta_\eta}{d\eta_+}\right) \tag{A.3}$$

which we have added and/or propagated to the other errors (due to the large size of our statistical samples the two purely statistical contributions to $\delta_x$ turn out to be essentially negligible compared to the others).

Acknowledgements: This work has been supported by Rutgers University, by CNR–GNMF, and it is part of the research program of the European Network on: "Stability and Universality in Classical Mechanics", # ERBCHRXCT940460,